\documentclass[apj,numberedappendix]{emulateapj}
\usepackage{apjfonts}
\shorttitle{ICM ENTROPY FROM X-RAY AND SZ DATA}
\shortauthors{CAVALIERE, LAPI \& REPHAELI}
\journalinfo{Accepted by The Astrophysical Journal}
\begin{document}
\title{Intracluster Entropy from Joint X-ray and Sunyaev-Zel'dovich Observations}
%
\author{A. Cavaliere\altaffilmark{1}, A. Lapi\altaffilmark{1, 2},
and Y. Rephaeli\altaffilmark{3}}
\altaffiltext{1}{Dip. Fisica, Universit\'a Tor Vergata, via Ricerca
Scientifica 1, I-00133 Roma, Italy} \altaffiltext{2}{Astrophysics
Sector, SISSA/ISAS, Via Beirut 2-4, I-34014 Trieste, Italy}
\altaffiltext{3}{School of Physics and Astronomy, Tel Aviv
University, 69978 Tel Aviv, Israel}
%
\begin{abstract}
The temperature and density of the hot diffuse medium pervading
galaxy groups and clusters combine into one significant quantity,
the entropy. Here we express the entropy levels and profiles in
model-independent forms by joining two observables, the X-ray
luminosity and the change in the CMB intensity due to the
Sunyaev-Zel'dovich (SZ) effect. Thus we present both global
\emph{scaling} relations for the entropy levels from clusters and
groups, and a simple expression yielding the entropy \emph{profiles}
in individual clusters from resolved X-ray surface brightness and SZ
spatial distributions. We propose that our approach provides two
useful tools for comparing large data samples with models, in order
to probe the processes that govern the thermal state of the hot
intracluster medium. The feasibility of using such a diagnostic for
the entropy is quantitatively assessed, based on current X-ray and
upcoming SZ measurements.
\end{abstract}
\keywords{cosmic microwave background - X rays: galaxies: clusters}
%
\section{Introduction}
%
Recent X-ray observations of the hot intracluster medium (ICM)
filling galaxy groups and clusters have indicated problems with both
the levels of the density, $n \sim 10^{-3}$ cm$^{-3}$, and the
distribution of the temperature $T$, with average values $kT\sim 10
-1/2$ keV from very rich clusters to poor groups. First, the lower
densities found in poor clusters and groups imply mass ratios $m/M$
of ICM to dark matter (DM) considerably below the cosmic baryonic
fraction approached in rich clusters (see Sanderson \& Ponman 2003,
Pratt \& Arnaud 2003).

Second, radial profiles $T(r)$ have been measured, if coarsely, in
several nearby clusters; they show a roughly isothermal plateau
extending out to substantial radii $r \approx 0.1-0.2\, R$ in terms
of the virial radius $R$, and then a decline by a factor $1/2$ out
to $R/2$ (Markevitch et al. 1998, De Grandi \& Molendi 2002,
Vikhlinin et al. 2005). Although still debated in detail, this
behavior differs from the predictions of simple modeling in terms of
a polytropic equation of state $T(r)\propto n^{\Gamma-1}$ with
constant index $1\leq \Gamma\leq 5/3$. It also differs at $r\la
0.2\, R$ from the outcomes of most state-of-the-art numerical
simulations, where $T(r)$ keeps rising toward the cluster center to
$r\sim 0.05\, R$ as discussed, e.g., by Borgani et al. (2004). A
central, limited dip of $T(r)$ observed in many clusters (e.g.,
Piffaretti et al. 2005) and often referred to as a ``cool core''
(Molendi \& Pizzolato 2001) is arguably attributed to a quenched
``cooling flow'' (see the discussion by Fabian 2004). Such dips
involve ICM fractions of some $10^{-3}$ and constitute a specific
issue not of direct concern here.

To focus the problems in the \emph{bulk} of the ICM, temperature and
density are conveniently combined into a significant single
quantity, the specific entropy $\mathfrak{s}$; here we adopt the
widely used adiabat $K\equiv kT\, n^{-2/3}$, which is related to
$\mathfrak{s}$ by $K\propto e^{2\, \mathfrak{s}/3\, k}$ (Bower 1997;
Balogh, Babul \& Patton 1999). The quantity $K$ constitutes the
simplest combination of $n$ and $T$ that is invariant under
adiabatic processes in the ICM; nonadiabatic processes compete to
produce different levels and profiles of $K$, that cooling always
lowers and various heating processes tend to raise.

When the data on $n$ and $T$ are combined to yield $K$, it is found
that (i) in moving from clusters to groups the \emph{levels} of $K$
decline only slowly (in fact, like $K\propto T^{2/3}$) taking on
values from around $10^{3}$ to some $10^{2}$ keV cm$^2$; (ii) in
clusters the entropy \emph{profiles} $K(r)$ have to be described in
terms of a running index
\begin{equation}
\Gamma(r) \equiv {5\over 3} + {d\ln{K}\over d\ln n}~,
\end{equation}
that implies profiles $K(r)$ flattening inwards.

These findings may be unified under the heading of an entropy
``\emph{excess}'' in the ICM. The baseline is provided by
gravitational heating, the basic process that affects the thermal
state of the ICM during its inflow into the structures as they are
built up by standard hierarchical clustering (Peebles 1993). During
the buildup, the DM component sets the gravitational potential wells
into which the baryons fall, starting with the cosmic density ratio
close to $0.16$ (Bennett at al. 2003). If the baryons start cold,
they fall in supersonically, and are shock-heated to temperatures
$T$ close to the virial value $T_V$ (Cavaliere, Menci \& Tozzi 1999;
Tozzi \& Norman 2001; Voit et al. 2003). However, such a process
produces (see Ponman, Sanderson, \& Finoguenov 2003): (i) entropy
levels lower than observed in groups and poor clusters; (ii) entropy
profiles in rich clusters that decline too steeply inwards, in fact,
as $K(r)\propto r^{1.1}$.

Such drawbacks have raised a wide debate concerning the nature of
the additional processes that increase the entropy of the ICM;
suggestions include the following possibilities. Energy drain by
radiative cooling extended well beyond the very central cool core
may cause much of the cold gas to condense into stars, leaving a
residual diffuse component with higher entropy (Voit \& Bryan 2001);
however, such a cooling alone would produce too much cold gas and
too many stars (e.g., McCarthy et al. 2004). Repeated energy inputs
by supernovae and active galactic nuclei preheat the gas external to
forming groups or clusters, raise its state to a higher adiabat, and
are expected to hinder its inflow under any model (Evrard \& Henry
1991; Kaiser 1991); specifically, as the inflow becomes less
supersonic, the shocks are weakened while additional entropy is
carried in (Valageas \& Silk 1999; Wu, Fabian \& Nulsen 2000;
Cavaliere, Lapi \& Menci 2002a). Energy impulsively discharged by
powerful quasars into their ambient medium propagates outwards
beyond the host galaxy into a surrounding group (Mazzotta et al.
2004a; McNamara et al. 2005); the propagation may occur in the form
of outgoing, moderately supersonic blast waves which at the leading
shocks raise the entropy in the gas they sweep up (Lapi, Cavaliere
\& Menci 2005). Finally, in rich clusters thermal or turbulent
diffusion may be effective in depositing and/or smearing out entropy
inward of $r\approx R$ (Narayan \& Medvedev 2001; Kim \& Narayan
2003; Lapi \& Cavaliere, in preparation).

Pinning down the leading process clearly requires precise
measurements of the levels and profiles of the all-important adiabat
$K$. The state variables $n$ and $T$ that determine $K$ may be
measured with X-ray observations (see Sarazin 1988). The bolometric
bremsstrahlung emissions $L_X \propto n^2\, \sqrt{T}\, R^3\approx
10^{42} - 10^{45}$ erg s$^{-1}$ primarily provides $n$, while
spatially resolved spectroscopy (continuum plus emission lines) can
yield the temperatures $T$ when enough photons are collected.

On the other hand, the hot ICM electrons inverse-Compton scatter the
CMB photons crossing a cluster; this Sunyaev-Zel'dovich effect (SZ,
Sunyaev \& Zel'dovich 1972) constitutes a major ICM (and
cosmological) probe in the $\mu$wave and submm bands, as reviewed by
Rephaeli (1995), Birkinshaw (1999), and Carlstrom, Holder \& Reese
(2002). The intensity of the thermal SZ effect is given by the
Comptonization parameter $y\propto n\,T\,R$, which constitutes a
good measure of the electron pressure $p\propto n\, T$ in the ICM;
so far it has been observed in many rich clusters at the expected
levels $y\approx 10^{-4}$ (Zhang \& Wu 2000; Grainge et al. 2002;
Reese et al. 2002; Benson et al. 2004). The main dependencies of
$L_X\propto n^2$ and $y\propto p$ strongly motivate us to employ the
results of X-ray and SZ observations combined as to provide the
adiabat in the form $K\propto p\, n^{-5/3}$, in order to better
probe properties of, and processes in the ICM.
%
\section{Scaling the entropy levels from clusters to groups}
%
The levels $K_{0.1}$ of the entropy (usually sampled at $r\approx
0.1\, R_{200}\approx 0.08\, R$, see Ponman et al. 2003) and of the
X-ray luminosity $L_X$ (dominated by the inner ICM anyway) are
linked by the relation
\begin{equation}
K_{0.1}/ K_g\propto (L_X/ L_g)^{-1/3} (T/ T_V)^{7/6}~,
\end{equation}
derived in Appendix A. In moving from clusters to groups, this
highlights the deviations from the quantities $K_g\propto H^{-4/3}\,
T_V$ and $L_g\propto H\, T_V^2$ that provide the baseline scaling --
including the Hubble parameter $H(z)$ -- set by the pure
gravitational heating (see \S~1); the latter implies $T\approx T_V$,
and baryon densities simply proportional to those of the DM (Kaiser
1986).

Eq.~(2) has been first proposed by Cavaliere, Lapi \& Menci (2002b);
similar expressions have been subsequently used by Dos Santos \&
Dor\`e (2002), and McCarthy et al. (2003), in the context of
specific models for entropy enhancement. Here we stress the
\emph{model-independent} nature of Eq.~(2); the latter holds
whenever the ICM fills in a statistical, virial-type equilibrium a
DM potential well, and depends weakly on specific DM distributions
or specific equations of state for the ICM. This is demonstrated in
Appendix A and Table~A1, where we show that from clusters to groups
the prefactor on the r.h.s. of Eq.~(2) depends weakly on $T$, so as
to closely preserve the above scaling.

We illustrate the role that Eq.~(2) may play in probing the
processes that set the ICM thermal state. We first recall that in
poor clusters and groups with $kT$ ranging from about $4$ to about
$1/2$ keV the observed values of $L_X$ are found to be progressively
lower than $L_g$ by factors from $10^{-1}$ to $10^{-3}$, with a
steep decline close to $L_X \propto T^3$ (Osmond \& Ponman 2004). In
other words, moving toward smaller and cooler structures the ICM is
found to be progressively underluminous, hence \emph{underdense}
relative to the gravitational values. The outcome is even more
surprising on two accounts: smaller structures ought to have
condensed earlier from a denser Universe; moreover, emission lines
from highly excited metals add to the bremsstrahlung continuum to
yield a flatter gravitational scaling $L_X \approx L_g \propto T$
for $kT < 2$ keV (see Appendix A). According to Eq.~(2), the
systematically \emph{lower} values of $L_X/L_g$ will correspond to
\emph{higher} entropy levels, exceeding $K_g$ by factors up to $10$.
In detail, the steep relation $L_{X} \propto T^3$ will reflect into
a flat behavior $K_{0.1}\propto T^{2/3}$, which indeed agrees with
recent data analyses (see Ponman et al. 2003).

On the other hand, a number of authors (e.g., Mahdavi et al. 1997;
Roussel, Sadat \& Blanchard 2000) have questioned the significance
of the data concerning groups, given various observational and
systematic uncertainties related to the limited statistics and to
the necessary subtraction of considerable contributions from single
galaxies. Recent critical assessments of the data (Mushotzky 2004;
Osmond \& Ponman 2004) acknowledge the relevance of these problems
in individual groups, but confirm the general trend toward lower
densities in poorer systems, albeit with a wide scatter.

To cross-check these low X-ray luminosities $L_X$, one may
preliminarily use the SZ effect; in fact, these two independent
probes of the ICM density are expected to correlate according to
\begin{equation}
y_0/ y_g\propto (L_X/ L_g)^{1/2} (T/ T_V)^{3/4}~.
\end{equation}
Here $y_0$ is the Comptonization parameter integrated along a
``central'' line of sight (l.o.s.), and $y_g\propto H\, T_V^{3/2}$
gives its gravitational scaling after Cole \& Kaiser (1988); see
Appendix A for the derivation and added detail, including a
discussion of the contributions to $y_0$ from the outer ICM. When
$L_X\propto T^3$ applies, Eq.~(3) predicts the scaling $y_0\propto
T^2$ to hold in equilibrium conditions (Cavaliere \& Menci 2001).
Such systematically lower values of $y_0 /y_g \propto T^{1/2}$ may
be tested or bounded with high-sensitivity SZ observations of poor
clusters and groups, as initiated by Benson et al. (2004).

Then one can proceed to directly obtain the entropy levels from the
equilibrium relations Eqs.~(2) and (3). Upcoming SZ measurements
will provide high-sensitivity data, which will be conveniently
joined with X-ray fluxes to obtain entropy levels that scale as
\begin{equation}
K_{0.1}\propto H^{-8/3}\, y_0^{10/3}\, L_X^{-2}~,
\end{equation}
in equilibrium conditions but otherwise in a model-independent way.
In moving to groups where line emission is important, Eq.~(4) goes
over to $K_{0.1}\propto H^{-5/3}\, y_0^{4/3}\, L_X^{-1}$; Appendix A
gives the details, including the slow dependence on $T$ of the
prefactor. Note that the present approach bypasses two difficulties
(discussed, e.g., by Mazzotta et al. 2004b) in deriving the ICM
temperatures from spatially resolved X-ray spectroscopy: (i) the
latter requires more photons than imaging, so with comparable
observation times and efficiencies, precise $T$ are obtained at
smaller radii than $n$; ii) in a multiphase medium spectroscopic
determinations of $T$ are biased toward the colder regions.

The integrated SZ flux $Y\propto y_0\,R^2/d_A^2$, where $d_A(z)$ is
the angular diameter distance, may be used instead of the central
parameter $y_0$; in terms of $Y$, Eq.~(4) may be recast as
\begin{equation}
K_{0.1}\propto d_A^{20/9}\, H^{2/3}\, Y^{10/9}\, L_X^{-8/9}~.
\end{equation}
Using this relation has two advantages over Eq.~(4). Not only is $Y$
a more robust observable in clusters with a cool core (e.g., Benson
et al. 2004), but also $Y\propto n\, T\, R^3$ has the same $R^3$
dependence as $L_X$. So the combination is largely free of the size
effects due to the scaling $R\propto T^{1/2}$ for coeval structures
(cf. Eq.~[A2]); it rather highlights the dependence on the baryonic
fraction.

So far we have dealt with the overall entropy \emph{levels}
determined from the integrated X-ray and SZ observables. But the
entropy \emph{profiles} of individual clusters carry additional
information; clearly this will be difficult to extract from the
prefactors in global relations like Eqs.~(2)-(5), encased as it is
under integrated forms within ratios or products of shape factors,
as made explicit in Appendix A.
%
\section{Resolving the entropy profiles in clusters}
%
To go beyond the above global scaling laws, spatially resolved X-ray
and SZ observations are clearly required. In X rays,
\textit{Chandra} and \textit{XMM-Newton} are already providing data
at a few arcsec resolutions; but obtaining both the good spatial and
\emph{spectral} resolution needed to measure $T(r)$ and derive
$K(r)$ still presents a challenge, especially in the outer ICM of
distant clusters. Improvements will again be provided by joining
measurements of X-ray brightness and of SZ effect with comparable
levels of spatial resolutions.

In fact, the prospective good resolution and high sensitivity (added
to multi-frequency capabilities) of the many upcoming SZ experiments
(for a recent review, see Birkinshaw 2004) stimulate renewed impetus
for considering the deprojected quantities that have been discussed
from time to time (e.g., Silk \& White 1978, Cavaliere 1980).

Here our specific proposal is to use joint X-ray and SZ measurements
to derive directly from the data the radial entropy profiles $K(r)$
in the ICM. We will work in terms of the X-ray surface brightness
and of SZ intensity distributions, integrated along a l.o.s at a
projected distance $s$ from the cluster center. For spherically
averaged ICM distributions these observables are related to the
corresponding volume quantities by
\begin{eqnarray}
\nonumber\\
\nonumber \ell_X (s) &=& {\Lambda_0\over 2\pi\,(1+z)^4}\, \int_s^{R}{dr}\, {r\over \sqrt{r^2-s^2}}\, n^2(r)\, \sqrt{T(r)}\\
\\
\nonumber y (s) &=& {2\, k\sigma_T\over m_e c^2}\, \int_s^{R}{dr}\, {r\over \sqrt{r^2-s^2}}\, n(r)\, T(r)~,\\
\nonumber
\end{eqnarray}
where the constants are defined after Eq.~(A1). Note how strongly
the X-ray surface brightness decreases as the redshift $z$
increases, while the SZ effect is $z$-independent, apart from the
dilution effect related to the size of the instrumental beam.

To construct the adiabat $K= kT\, n^{-2/3}$ we need to retrieve the
volume quantities $n(r)$ and $T(r)$ from the projected quantities
$\ell_X (s)$ and $y(s)$ given by Eqs.~(6). Actually, the latter are
in the form of two Abel integral equations, which may be solved for
the volume quantities along the lines recalled in Appendix B; the
results come to
\begin{eqnarray}
\nonumber\\
\nonumber n^2\, \sqrt{kT} &=& {4\, k^{1/2}\, (1+z)^4\over \Lambda_0}~~ {1\over r}\, {d\over dr}\, \int_{R}^{r}{ds}\, {s\over \sqrt{s^2-r^2}}\, \ell_X (s)\equiv \mathcal{L}_X (r)\\
\\
\nonumber n\, kT &=& {m_e c^2\over \pi\sigma_T}~~ {1\over r}\, {d\over dr}\, \int_{R}^{r}{ds}\, {s\over \sqrt{s^2-r^2}}\, y(s)\equiv \mathcal{Y} (r)~. \\
\nonumber
\end{eqnarray}
We stress that the two volume quantities $\mathcal{L}_X(r)$ and
$\mathcal{Y}(r)$ are defined directly in terms of the observed
surface or l.o.s. observables $\ell_X (s) $ or $y(s)$.

Eqs.~(7) are easily combined (over the spatial range common to both
data sets) to yield the entropy profile in the simple,
\emph{model-independent} form
\begin{equation}
K(r) = \mathcal{Y}^{14/9}(r)\, \mathcal{L}_X^{-10/9}(r)~.
\end{equation}
So far we have not assumed any specific equation of state for the
ICM, rather we derive it from $K(r)$ above; in particular, the local
polytropic index defined in Eq.~(1) reads
\begin{equation}
\Gamma(r) = 3\, \left[2\, {d\ln\mathcal{L}_X\over d\ln
\mathcal{Y}}-1\right]^{-1}~.
\end{equation}

To illustrate the importance of the profile $K(r)$ we contrast two
physical conditions. On the one hand, a reference baseline is
provided once again by the gravitational heating associated with
smooth accretion of successive shells (involving DM plus gas, as
started by Bertschinger 1985) during the hierarchical buildup of a
rich cluster. This ``surface'' process taking place at the outskirts
of the growing structure yields $K(r)\propto r^{1.1}$ and
$\Gamma\approx 1.1$ in the region $r\approx R$ (Tozzi \& Norman
2001). On the other hand, the actual inflow is likely to include
additional ``volume'' processes induced by infall and sinking of
lumps, the well-known merging component in the process of structure
formation (see Norman 2005). These dynamical events can drive
subsonic turbulent diffusion and related entropy transport from the
outskirts toward the inner regions (see Goldman \& Rephaeli 1991;
Ricker \& Sarazin 2001; Inogamov \& Sunyaev 2003); clearly these
processes will enforce entropy profiles flatter than $K(r)\propto
r^{1.1}$, a matter expanded upon elsewhere (Lapi \& Cavaliere, in
preparation). So it is seen that the entropy profiles can help
unveiling the actual physical conditions in clusters.

\begin{figure*}[t]
\epsscale{.9}\plotone{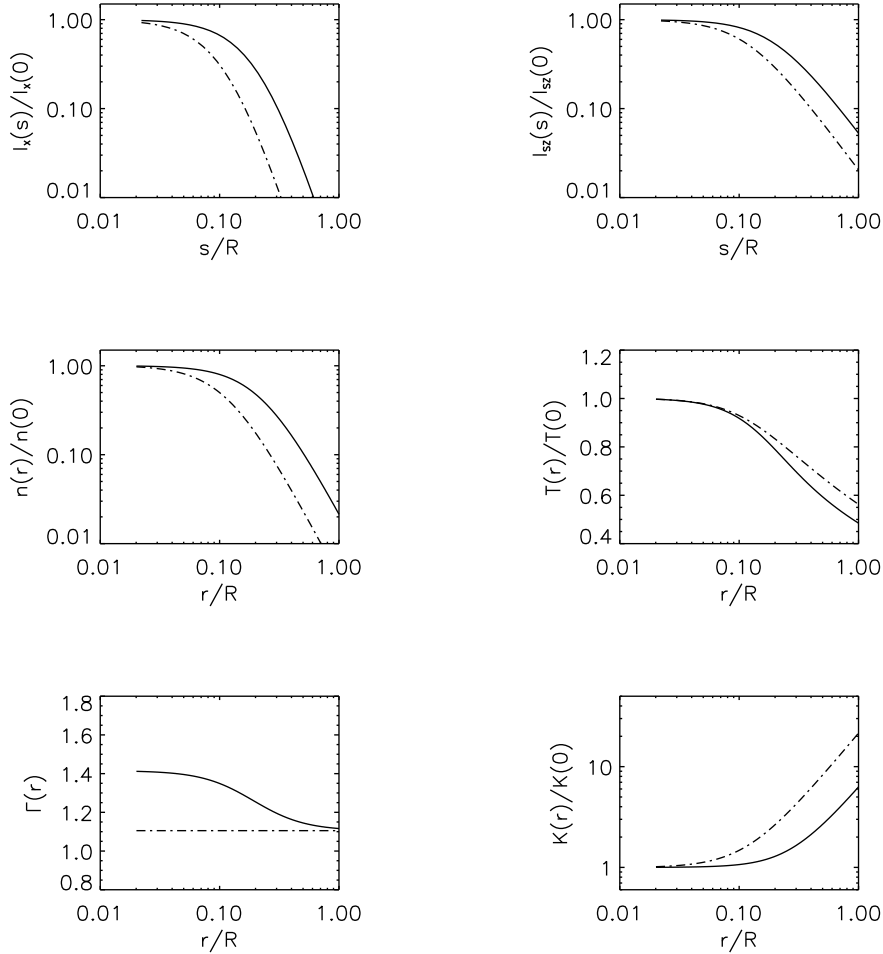}\caption{Input profiles for the
mock experiments described in \S~4: X-ray surface brightness, SZ
intensity distribution, and radial profiles of density, temperature,
entropy and polytropic index. \textit{Dot-dashed} lines are for the
simple polytropic model $\mathcal{A}$ with $\Gamma=1.1$, while
\textit{solid} lines are for the nonpolytropic model $\mathcal{B}$
with running index $\Gamma(r)$; see main text for details.}
\end{figure*}

\begin{figure*}[t]
\epsscale{.9}\plotone{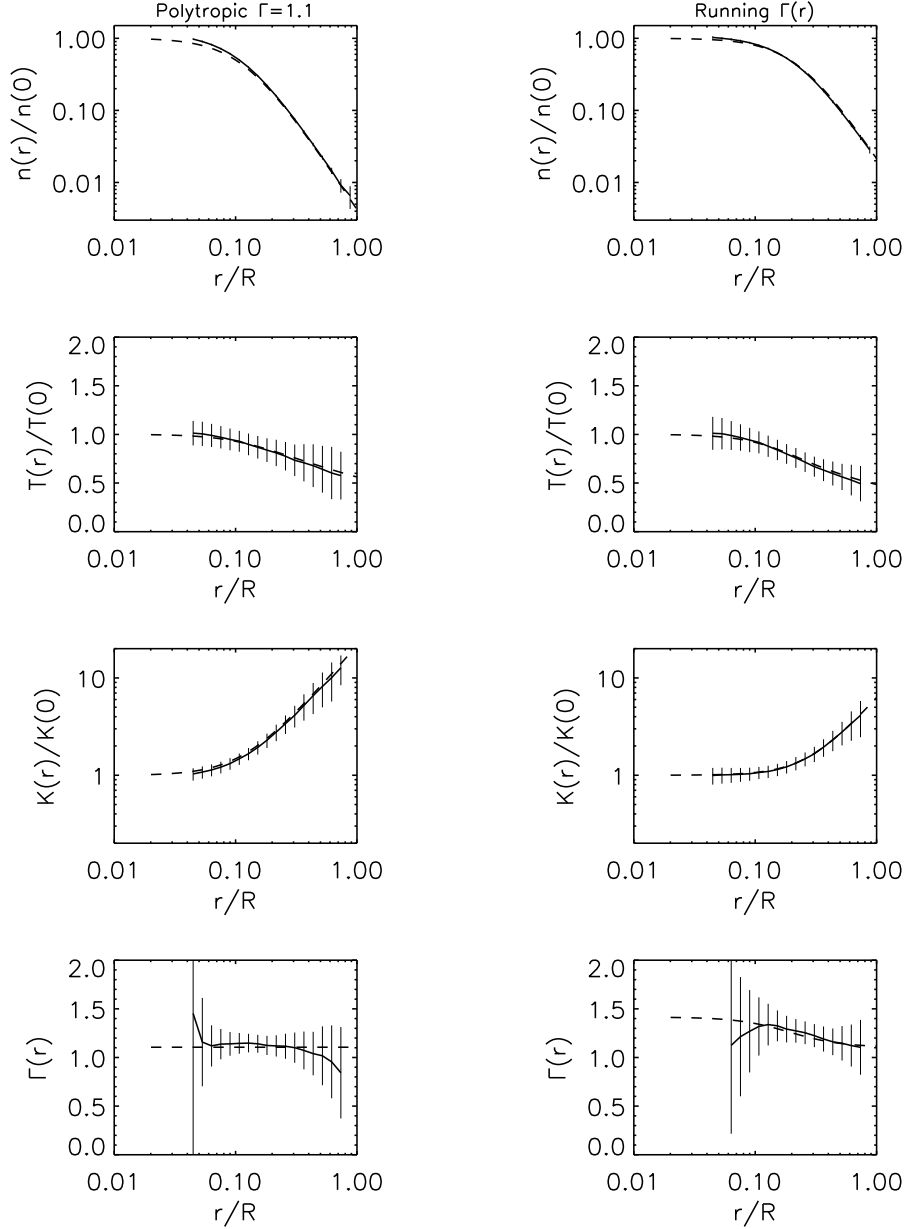}\caption{Reconstructed profiles
from the the mock experiments of \S~4: radial profiles of density,
temperature, entropy, and polytropic index are shown from top to
bottom as \textit{solid} lines, with $1\sigma$ error bars;
\textit{dashed} lines show the input profiles. \textit{Left} column
refers to model $\mathcal{A}$ and \textit{right} column to model
$\mathcal{B}$.}
\end{figure*}

Note that Eqs.~(7) also yield the density and temperature profiles
in the form
\begin{equation}
n(r) = \mathcal{L}_X^{2/3}(r)\, \mathcal{Y}^{-1/3}(r)~~~~~~~~~kT(r) = \mathcal{Y}^{4/3}(r)\,
\mathcal{L}_X^{-2/3}(r)~.
\end{equation}
Correspondingly, the local cooling timescale for bremsstrahlung
emission reads
\begin{equation}
t_c(r) ={3\, k^{1/2}\over \Lambda_0}\, {\mathcal{Y}(r)\over \mathcal{L}_X(r)}~.
\end{equation}
Comparing this to the Hubble time provides a criterion to avoid the
inclusion of a region like a cool core, where the physical
conditions are very different from the rest of the cluster.

So far, our derivations have been free of specific assumptions
concerning models for the DM gravitational potentials, or
hydrostatic equilibrium, or equation of state. Assuming now detailed
hydrostatic equilibrium, one can use the ICM quantities to trace the
DM mass distribution (e.g., Bahcall \& Sarazin 1977; Fabricant \&
Gorenstein 1983), and obtain the result
\begin{equation}
M(<r) = {3\over 4\, G\, \mu m_p}\, {r^2\over
\mathcal{L}_X^{2/3}(r)}\, {d\over d r}\mathcal{Y}^{4/3}(r)~;
\end{equation}
here $\mu\approx 0.6$ is the mean molecular weight for a plasma of
one-third solar metallicity, $m_p$ is the proton mass and $G$ is the
gravitational constant. Equivalently, Eq.~(12) may be used to derive
a model for the DM gravitational potential, to be compared with
specific forms such as King's (1972) and NFW (Navarro, Frenk \&
White 1997), or with the data obtained from gravitational lensing
(e.g., Lombardi et al. 2005).
%
\section{Mock experiments}
%
Given that the imaging capabilities of the present X-ray instruments
(\textit{Chandra} and \textit{XMM-Newton}) have attained resolutions
down to about $1''$ and fractional sensitivities of order $10^{-3}$
or better, what can be learned right now by joining these X-ray data
with the current, considerably coarser SZ observations?

In fact, here problems arise even for the global scaling laws, and
more for profile reconstructions. As discussed in detail by Diaferio
et al. (2005), considerable discrepancies as to the scaling laws are
found among current SZ samples produced with different instruments,
even when the samples have many sources in common. On the other
hand, to effectively constrain a fit to the observed SZ
distributions often it has been found necessary to use also X-ray
information (see Reese et al. 2003, Benson et al. 2004); this breaks
the independence of these two observables that is important for
deriving $K(r)$.

A step forward has recently been taken by the observations of the
X-ray cluster RX J1347-1145 with the Nobeyama telescope (Komatsu et
al. 2001); the good resolution of order $15''$ obtained in these
measurements enabled deriving a truly independent SZ profile. In
analyzing these data Kitayama et al. (2004) excise the most
asymmetric sector, and spherically average the rest to deconvolve
the temperature profile, finding results consistent with - but still
no better than - the X-ray spectroscopy.

In the light of upcoming SZ capabilities of sub-arcmin resolution
and sensitivities better than $ 10 \,\mu$K, we consider next the
prospects from measurements with instruments or experiments such as
SZA (Holder et al. 2000), AMI (Jones 2002), AMiBA (Lo 2002), MITO
(Lamagna et al. 2005), OLIMPO (Masi et al. 2003), ACT (Kosowsky
2003), SPT (Carlstrom et al. 2003), APEX-SZ (Schwan et al. 2003),
CARMA (Woody et al. 2004). It is expected that many hundreds of
clusters and groups will be detected in a survey mode, with pointed
observations of many clusters to sensitivity levels of $10\, \mu$K
or better. In preselected areas, resolutions down to a few arcsecs
will be eventually provided by ALMA (see \url{http://
www.alma.nrao.edu/}), at frequencies on both sides of $220$ GHz, the
crossover point where the thermal SZ effect goes from negative to
positive.

We build mock renditions of the prospective observations as
described next. We make use of two \emph{independent}, flexible
parameterizations for the spatial distributions of the X-ray
brightness and SZ effect; both are of the form
\begin{equation}
I(s) = I(0)\, \left[1+\left({s\over a}\right)^2\right]^{-b}~
\end{equation}
often adopted for either one, though best suited to clusters with no
cool core. Here $a$ is the inner, characteristic radius and $b$ is
the shape parameter for the outer profile (e.g., Reese et al. 2002;
Schmidt, Allen \& Fabian 2004). Note that $a/\sqrt{b}$ provides an
effective ``core radius'', while $I(s)\propto s^{-2\, b}$
approximately holds for large $r$. To preserve the independence of
the two fits, we allow for \emph{different} values of the parameters
$a_X, a_{SZ}$ and $b_X, b_{SZ}$ in Eq.~(13); note that the
isothermal $\beta$-model (Cavaliere \& Fusco-Femiano 1976; Jones \&
Forman 1984) would obtain as the particular case where $a_X=a_{SZ}$,
$b_X= 3\, \beta-1/2\approx 1.5$ and $b_{SZ}= (3\, \beta-1)/2\approx
0.5$ apply.

First, let us disregard the limitations due to finite resolution and
precision; then the fitting formulae Eq.~(13) are easily
Abel-inverted, and lead to sharp volume distributions $n^2\,
T^{1/2} \propto [1+(r/a_X)^2]^{-b_X-1/2}$ and $n\, T \propto
[1+(r/a_{SZ})^2]^{-b_{SZ}-1/2}$ for the X-ray emissivity and SZ
effect, respectively. Using these in Eqs.~(8) and (9), the entropy
profile reads
\begin{equation}
{K(r)/K(0)}=\left[1+\left(r/ a_X\right)^2\right]^{5\,
(1+2\,b_X)/9}\, \left[1+\left({r/
a_{SZ}}\right)^2\right]^{-7\,(1+2\,b_{SZ})/9}~,
\end{equation}
while the polytropic index is given by
\begin{equation}
\Gamma(r) = 3\, \left[\frac{2+4\, b_{X}}{1+2\, b_{SZ}}\,
\frac{r^2+a_{SZ}^2}{r^2+a_X^2}-1\right]^{-1}~.
\end{equation}
These runs are plotted in Fig.~1 for the specific parameter values
given in the caption. Note that, when $a_X=a_{SZ}$ is set, a
constant value of $\Gamma = (3+6\, b_{SZ})/(4\, b_X-2\, b_{SZ}+1)$
would obtain. Then the bounds $(2\, b_X-1)/4\leq b_{SZ}\leq (5\,
b_X-1)/7$ apply; the upper limit corresponds to $\Gamma=5/3$, i.e.,
the isentropic profile, while the lower limit corresponds to
$\Gamma=1$, i.e., the isothermal profile.

Next we proceed to generate and analyze mock SZ experiments on
introducing prospective resolutions and precisions. We focus on two
input models: $\mathcal{A}$) a simple polytropic model with constant
index $\Gamma\approx 1.1$, which corresponds to Eq.~(13) with
$b_X=2.1$, $b_{SZ}=0.9$, $a_X/\sqrt{b_X}=0.08\, R$ and $a_X=a_{SZ}$;
$\mathcal{B}$) a nonpolytropic model based on Eq.~(13) with
$b_X=2.1$, $b_{SZ}=0.9$, $a_X/\sqrt{b_X}=0.15\, R$ and
$a_{SZ}/\sqrt{b_{SZ}}=0.21\, R$, implying a running index
$\Gamma(r)$. These input profiles are illustrated in Fig.~1.

Having in mind rich clusters at $z\la 0.1$, we bin the data into
logarithmically equal intervals limited by a central resolution of
$10''$; higher resolution, such as eventually attainable with ALMA,
will extend the limiting $z$ beyond $0.5$. Following Yoshikawa \&
Suto (1999; see also Tsutsumi et al. 2005, in preparation), we
assign $1\, \sigma$ uncertainties $\Delta I_X\approx 2\times
10^{-4}\, I_X(0)$ and $\Delta I_{SZ}\approx 10^{-2}\, I_{SZ}(0)$ to
the input X-ray surface brightness and SZ intensity distributions,
respectively. Using a standard Monte Carlo scheme we randomly sample
Gaussian-distributed values for $I_X$ and $I_{SZ}$, then perform the
reconstruction, and finally average over $1000$ realizations.

Fig.~2 shows for each model $\mathcal{A}$ and $\mathcal{B}$ the
input (\textit{dashed} lines) and reconstructed profiles
(\textit{solid} lines with $1\, \sigma$ error bars) of the ICM
density $n(r)$, temperature $T(r)$, entropy $K(r)$ and polytropic
index $\Gamma(r)$.

It is seen that the average reconstructed profiles are quite similar
to the input profiles over a considerable range, in spite of
uncertainties and of some systematic deviations increasingly induced
at small and large radii by our emulated resolution and sensitivity.
It is also seen that the uncertainties in $T(r)$, dominated by the
assumed limitations of the SZ observations, considerably exceed
those in $n(r)$, mainly contributed by the X-ray observations. On
the other hand, the former do not grow fast toward the dimmer
outskirts, unlike the X-ray observations particularly of distant
clusters in the presence of instrumental or astrophysical
background. We add that uncertainties and deviations reduce to about
one half when we assume levels of sensitivity and of spatial
resolution twice higher than the values used in Fig.~2.
%
\section{Discussion and conclusions}
%
To make optimal use of the precise and resolved, upcoming SZ data we
have proposed a simple formalism, which is model-independent even
though formulated here for spherically \emph{averaged} symmetry.
Different geometries may be considered along the lines discussed by
Binney \& Strimpel (1978), and advanced by Zaroubi et al. (2001).

We propose that SZ and X-ray measurements jointly provide useful
diagnostic tools aimed at assessing the thermal state of the ICM
with the use of large data sets. Specifically, Eqs.~(4) or (5)
provide simple, global scaling laws for the \emph{levels} of the
central entropy $K_{0.1}$ from clusters to groups; on the other
hand, in relatively nearby rich clusters Eqs.~(8) and (9) directly
yield the resolved \emph{profiles} $K(r)$ and the running polytropic
index $\Gamma(r)$. The former relations will provide insight on the
processes governing the overall thermal state of the ICM in clusters
and in groups. The latter relations will probe, e.g., the
effectiveness of thermal or turbulent diffusion in establishing the
local equation of state for the ICM within rich clusters.

We stress that the scaling in Eq.~(3) yields what actually
represents a lower limit to $y_0$, valid for the ICM in equilibrium
within the potential wells. As shown in Table~A1 and commented below
it in Appendix A, the contribution to $y_0$ from the outer range
$r\ga 0.2\, R$ is limited to $10-15\%$, and encased in the shape
factors; their slow dependence on system temperature can only yield
moderate deviations when the outer and inner regions are described
by the same monotonically decreasing hydrostatic distribution. By
the same token, upward deviations of $y_0$ will strongly indicate
contributions from outer nonequilibrium plasma, overpressured and
outflowing from the wells, an issue that discussions with P.
Mazzotta stimulated us to stress. In fact, Lapi, Cavaliere \& De
Zotti (2003, see their Fig.~3) have computed the transient upper
limits expected in $5-10\%$ early bright galaxies and surrounding
groups when powerful quasars flare up and drive outgoing blast waves
that sweep out part of the ambient medium while raising its pressure
and entropy.

In summary, SZ observations with sensitivities of $10\, \mu$K or
better and sub-arcmin resolutions will provide the means to scale
entropy \emph{levels} in poor clusters and groups out to $z\approx
1.5$. Resolutions of order $10''$ will yield pressure and entropy
\emph{profiles}, most valuable in the outskirts of relatively nearby
clusters as long advocated by Rephaeli (e.g., 2005). Higher
resolutions of a few arcsecs will enable reconstructing the entropy
profiles in rich clusters toward $z\approx 1$, where SZ may
effectively compete with X rays being unaffected by cosmological
dimming. These SZ effect frontiers will afford unprecedented insight
on the astrophysics of the ICM.
%
\begin{acknowledgements}
We thank our referee for several helpful comments and suggestions.
We are indebted to P. Mazzotta for critical reading of an earlier
version of this paper, and for insightful discussions of the X-ray
and SZ differential sensitivities. Work supported by grants from
INAF and MIUR at the University of Rome Tor Vergata, and from the
Israel Science Foundation at Tel Aviv University.
\end{acknowledgements}
%
\begin{appendix}
%
\section{The scaling laws derived}
%
The integrated X-ray luminosity and the ``central'' SZ parameter (in
the non-relativistic case) are given by
\begin{eqnarray}
\nonumber\\
\nonumber L_X &=& 4\pi\, \Lambda_0\, n_2^2\, T_2^{\epsilon}\, R^3\, \int_0^1{dx}\,x^2\, \left({n\over n_2}\right)^2\, \left(T\over T_2\right)^{\epsilon}\\
\\
\nonumber y_0 &=& {2\, k \sigma_T\over m_e c^2}\, n_2\, T_2\, R\, \int_0^1{dx}\, {n\over n_2}\, {T\over T_2}~,\\
\nonumber
\end{eqnarray}
where we have used averaged spherical symmetry in terms of the
non-dimensional radial coordinate $x=r/R$, and have normalized the
electron density $n$ and the temperature $T$ to their values $n_2$
and $T_2$ at $r=R$. Here the expression $\Lambda_0\, T^{\epsilon}$
approximates the X-ray emissivity in terms of the normalization
$\Lambda_0$ (taking on values around $2\times 10^{-27}$ erg cm$^3$
s$^{-1}$ K$^{-1/2}$ for bremsstrahlung emission) and of the
power-law index $\epsilon$ (ranging along the cooling curve from
$\epsilon\approx 1/2$ for bremsstrahlung-dominated emission to
$\epsilon \approx -1/2$ for important line emissions at $kT < 2$
keV; see Sutherland \& Dopita 1993).

The statistical equilibrium of the ICM within a DM potential well of
depth marked by the virial temperature $T_V$ implies the size
dependence
\begin{equation}
T_V \propto H^{2}\, R^{2}~.
\end{equation}
Here $H(z)\equiv [\Omega_M\,(1+z)^3+\Omega_{\Lambda}]^{1/2}$ is
defined in terms of the virialization redshift $z$, and of the
present density $\Omega_M\approx 0.27$ for the non-relativistic
matter (DM $+$ baryons) and $\Omega_{\Lambda}\approx 0.73$ for the
dark energy in the Concordance Cosmology (Bennett et al. 2003); the
approximation $H(z)\propto (1+z)$ holds for $z < 1$.

Thus we obtain the scaling laws
\begin{equation}
L_X \propto \mathcal{S}_X\, H^{-3}\, (T_V/T_2)^{3/2}\, n_2^2\,
T_2^{\epsilon+3/2}~~~~~~~~y_0 \propto \mathcal{S}_y\, H^{-1}\,
(T_V/T_2)^{1/2}\, n_2\, T_2^{3/2}~,
\end{equation}
in terms of the two shape factors $\mathcal{S}_X$, $\mathcal{S}_y$
defined by the integrals appearing in Eqs.~(A1). As to the adiabat,
its levels $K_{0.1}$ at $r\approx 0.1\, R$ may be written as
\begin{equation}
K_{0.1}\propto \mathcal{S}_K\, {T_2\over n_2^{2/3}}
\end{equation}
in terms of the other shape factor $\mathcal{S}_K$. We will see
below that all factors: $T_2/T_V$, $\mathcal{S}_X$, $\mathcal{S}_y$,
and $\mathcal{S}_K$, vary only weakly from clusters to groups, and
so constitute minor corrections to the explicit scaling laws above.
This stems from the circumstance that the structural function
$n(r)/n_2$ entering the shape factors depends weakly on $T_2$ from
clusters to groups, as observed by Sanderson \& Ponman (2003) and
Pratt \& Arnaud (2003). Weak dependence of $n(r)$ is also expected
from all equilibrium models (as discussed in depth by Cavaliere et
al. 2002b); in fact, these show a weaker and weaker dependence in
passing from King (1972) to NFW gravitational potentials (due to
their DM concentration rising from clusters to groups, see Navarro,
Frenk, \& White 1997), or from isothermal to polytropic ICM (due to
a lesser density rise toward the center).

On eliminating $n_2$ from Eqs.~(A3) and (A4) we obtain the relations
\begin{equation}
y_0/y_g \propto (T_V/T_2)^{-1+\epsilon/2}\,
\left(L_X/L_g\right)^{1/2}~~~~~~~~K_{0.1}/ K_g \propto
(T_V/T_2)^{-1-\epsilon/3}\, \left(L_X/L_g\right)^{-1/3}~,
\end{equation}
which for $\epsilon=1/2$ yield Eqs.~(2) and (3) of the main text. We
have normalized $K_{0.1}$, $y_0$ and $L_X$ to the respective
gravitational scaling $K_g\propto H^{-4/3}\, T_V$, $y_g\propto H\,
T_V^{3/2}$, $L_g\propto H\, T_V^{\epsilon+3/2}$. Recall from \S~2
that these obtain on considering $T_2\approx T_V$, and $n_2\propto
H^2$ following the DM density, in turn proportional to the
background's; note that when line emission is important
$\epsilon=-1/2$ and $L_g\propto T_V$ apply, as stated in \S~2.

Moreover, on eliminating $n_2$ and the main $T_2$ dependence from
Eqs.~(A3) we find
\begin{equation}
K_{0.1}\propto H^{-(18-4\, \epsilon )/(9-6\, \epsilon )}\,
y_0^{(18+4\, \epsilon)/(9-6\, \epsilon )}\, L_X ^{-4/(3-2\,
\epsilon)}~,
\end{equation}
which for $\epsilon=1/2$ yields Eq.~(4) of the main text; for
simplicity we have omitted the slowly $T$-dependent prefactors. In
full, these read $(T_V/T_2)^{-1+\epsilon/2}\, \mathcal{S}_y\,
\mathcal{S}_X^{-1/2}$, $(T_V/T_2)^{-1-\epsilon/3}\, \mathcal{S}_K\,
\mathcal{S}_X^{1/3}$ and $(T_V/T_2)^{(9-2\, \epsilon )/(9-6\,
\epsilon)}\, \mathcal{S}_X^{4/(3-2\, \epsilon)}\,
\mathcal{S}_y^{-(18+4\, \epsilon)/(9-6\, \epsilon )}\,
\mathcal{S}_K$ on the r.h.s. of Eqs.~(A5) and (A6), respectively.

For the purpose of illustrating the slow dependence of $T_V/T_2$,
$\mathcal{S}_X$, $\mathcal{S}_y$, and $\mathcal{S}_K$ we now assume
a specific DM potential well, and detailed hydrostatic equilibrium
of the ICM with a specific relation between $n(r)$ and $T(r)$. For
the latter we adopt a polytropic distribution with $T(r)\propto
n(r)^{\Gamma-1}$ in terms of an average, constant index
$\Gamma\approx 1.1 - 1.3$; then the temperature run writes $T(r)/T_2
= 1+(\Gamma-1)\, \beta\, \Delta \phi(r)/\Gamma$, in terms of the
potential difference $\Delta \phi(r)=[\Phi(R)-\Phi(r)]/\sigma^2$
normalized to the DM one-dimensional velocity dispersion $\sigma$.
As to the DM potential $\Phi(r)$, we use the simple King form $\Phi
= -9\, \sigma^2\, \ln[(r/r_c)+(1+r^2/r^2_c)^{1/2}] \, r_c/r$ with
$r_c=R/12$, but similar or better results obtain on using the NFW
model. The parameter $\beta\equiv \mu m_p\, \sigma^2/k T_2=T_V/T_2$
encodes the $T_2$ dependence of the normalized profiles; it takes on
values varying in the range from $0.7$ in rich clusters to $0.5$ in
poor groups.

In Table~A1 we show our results; for simplicity $\epsilon=1/2$ is
adopted, but we add that $S_X$ changes less than $15\%$ for
$\epsilon=-1/2$ (e.g., for $\Gamma=1.2$ the value decreases from
$0.73$ to $0.65$ in groups with $\beta=0.5$). To put these changes
in context, recall from \S~2 that in moving from rich clusters with
$kT\approx 10$ keV to groups with $kT\approx 1$ keV, $L_X\propto
T^{3}$ is observed to decrease by factors $10^2$ or more,
$K_{0.1}\propto T^{2/3}$ by at least a factor $5$, and $y_0\propto
T^{2}$ is expected to decrease by $30$ or more; thus the $T$
dependencies of the shape factors in the scaling relations
constitute minor corrections. Actually, even slower variations occur
in the prefactors of Eqs.~(A5) and (A6) as they contain combinations
of shape factors whose variations tend to compensate; e.g., for
$\Gamma = 1.2$ the prefactors in Eqs.~(A5) vary by $0.9$ and $1.5$,
respectively, from clusters to groups.

As to sensitivity of the SZ effect to the outer wings of the ICM
distribution, note the bounds to their contribution set by the
observed outer decline of the temperature $T(r)$; to give an
example, the contribution to $y_0$ from the outer range $r\ga 0.2\,
R$ comes to $15-10\%$ when the polytropic index ranges between
$\Gamma\approx 1.1-1.3$. The $T$-dependence from clusters to groups,
the matter relevant to the scaling relations Eqs.~(3)-(5), is
bounded by the slow variation of the overall shape factor
$\mathcal{S}_y$ given in Table~A1.

\begin{deluxetable}{lcccccccccccc}
\tabletypesize{} \tablecaption{Shape factors, dependence on $T_2$}
\tablewidth{0pt} \tablehead{\colhead{} &
\multicolumn{3}{c}{$\Gamma=1.1$} & & \multicolumn{3}{c}{$\Gamma=1.2$} & & \multicolumn{3}{c}{$\Gamma=1.3$}\\
\colhead{$\beta$} & \colhead{$\mathcal{S}_X$} &
\colhead{$\mathcal{S}_y$} & \colhead{$\mathcal{S}_K$} & &
\colhead{$\mathcal{S}_X$} & \colhead{$\mathcal{S}_y$} &
\colhead{$\mathcal{S}_K$} & & \colhead{$\mathcal{S}_X$} &
\colhead{$\mathcal{S}_y$} & \colhead{$\mathcal{S}_K$}}\startdata
0.7...... & 1.66 & 7.43 & 0.19 & & 1.17 & 5.78 & 0.33 & & 0.96 & 4.94 & 0.47\\
0.6...... & 1.15 & 5.29 & 0.24 & & 0.91 & 4.42 & 0.37 & & 0.79 & 3.93 & 0.51\\
0.5...... & 0.84 & 3.82 & 0.29 & & 0.73 & 3.37 & 0.43 & & 0.66 & 3.10 & 0.56\\
\enddata
\tablecomments{The shape factors are computed for a polytropic ICM
in equilibrium within a King's DM potential well. Recall that
$\beta\equiv T_V/T_2$ ranges from $0.7$ to $0.5$ in moving from rich
clusters to poor groups.}
\end{deluxetable}

%
\section{The Abel equation}
%
We recall that the Abel integral equation has the form
\begin{equation}
f(x)=\int_x^q {dt}\, {F(t)\over \sqrt{t-x}}~;
\end{equation}
here $f(x)$ is a known function, $q$ a constant (that may be large),
and $F(t)$ the unknown function. To derive the latter, multiply both
sides of Eq.~(B1) by $(x-\xi)^{-1/2}$ and integrate over $x$ in the
interval [$\xi$, $q$]:
\begin{equation}
\int_{\xi}^q{dx}\, {f(x)\over \sqrt{x-\xi}}=\int_{\xi}^q{dx\over
\sqrt{x-\xi}}\, \int_x^q{dt} {F(t)\over \sqrt{t-x}} = \int_\xi^q{dt}\, F(t)
\int_\xi^t{dx} {1\over \sqrt{(x-\xi)\, (t-x)}}~,
\end{equation}
where the second equality is obtained on exchanging the integration
order; the last integral is simply $\pi$. Finally, differentiate
both sides of Eq.~(B2) with respect to $\xi$ to obtain
\begin{equation}
F(\xi) = {1\over \pi}\, {d\over d\xi}\, \int_q^\xi{dx}\, {f(x)\over
\sqrt{x-\xi}}~,
\end{equation}
which is the Abel inversion formula.

Note that on differentiating with respect to $\xi$ (with due
attention paid to the singularity of the kernel), Eq.~(B3) may be
recast into the alternative form
\begin{equation}
F(\xi) = {1\over \pi}\, \int_q^\xi{dx}\, {f'(x)\over \sqrt{x-\xi}}
+{1\over \pi}\, {f(q)\over \sqrt{q-\xi}}~,
\end{equation}
where the prime denotes differentiation with respect to $x$. But,
apart from problems with the boundary term $f(q)/\pi\,
\sqrt{q-\xi}$, Eq.~(B4) is less useful in the context of the main
text since it involves differentiation of the data or of their fit,
with the related uncertainties (possibly enhanced by real clumpiness
in the data) as discussed in detail by Lucy (1974) and Yoshikawa \&
Suto (1999).

\end{appendix}

%

%
%

\begin{references}
%
\reference{}Bahcall, J.N., \& Sarazin, C.L. 1977, ApJ, 213, L99

\reference{}Balogh, M.L., Babul, A., \& Patton, D.R. 1999, MNRAS,
307, 463

\reference{}Bennett, C.L., et al. 2003, ApJS, 148, 1

\reference{}Benson, B.A., et al. 2004, ApJ, 617, 829

\reference{}Bertschinger, E. 1985, ApJS, 58, 1

\reference{}Binney, J., \& Strimpel, O. 1978, MNRAS, 185, 473

\reference{}Birkinshaw, M. 1999, Phys. Rev., 310, 97

\reference{}Birkinshaw, M. 2004, in \textit{Clusters of Galaxies:
Probes of Cosmological Structure and Galaxy Evolution}, eds. J.S.
Mulchaey, A. Dressler, and A. Oemler (Cambridge: Cambridge Univ.
Press), 162

\reference{}Borgani, S., et al. 2004, MNRAS, 348, 1078

\reference{}Bower, R.G. 1997, MNRAS, 288, 355

\reference{}Carlstrom, J.E., Holder, G.P., \& Reese, E.D. 2002, ARA\&A, 40, 643

\reference{}Carlstrom, J.E., \& The SPT Collaboration 2003,
Highlights of Astronomy Vol. 13, ed. O. Engvold (San Francisco:
ASP), in press

\reference{}Cavaliere, A., \& Fusco-Femiano, R. 1976, A\&A, 49,
137

\reference{}Cavaliere, A. 1980, in \textit{X-ray astronomy}
(Dordrecht: D. Reidel Publishing Co.), 217

\reference{}Cavaliere, A., Menci, N., \& Tozzi, P. 1999, MNRAS, 308,
599

\reference{}Cavaliere, A., \& Menci, N. 2001, MNRAS, 327, 488

\reference{}Cavaliere, A., Lapi, A., \& Menci, N. 2002a, ApJ, 581, L1

\reference{}------------------------------------------- 2002b, in
ASP Conf. Series 268, \textit{Tracing Cosmic Evolution with Galaxy
Clusters}, eds. S. Borgani, M. Mezzetti, and R. Valdarnini (San
Francisco: ASP), 240

\reference{}Cole, S., \& Kaiser, N. 1988, MNRAS, 233, 637

\reference{}De Grandi, S., \& Molendi, S. 2002, ApJ, 567, 163

\reference{}Diaferio, A., et al. 2005, MNRAS, 356, 1477

\reference{}Dos Santos, S., \& Dor\`e, O. 2002, A\&A, 383, 450

\reference{}Evrard, A.E., \& Henry, J.P. 1991, ApJ, 383, 95

\reference{}Fabian, A.C. 2004, in AIP Conf. Ser. 703,
\textit{Plasmas in the Laboratory and in the Universe: New Insights
and New Challenges}, eds. G. Bertin, D. Farina, and R. Pozzoli
(AIP), 337

\reference{}Fabricant, D., \& Gorenstein, P. 1983, ApJ, 267, 535

\reference{}Goldman, I., \& Rephaeli, Y. 1991, ApJ, 380, 344

\reference{}Grainge, K., Jones, M.E., Pooley, G., Saunders, R., Edge, A.,
Grainger, W.F., \& Kneissl, R. 2002, MNRAS, 333, 318

\reference{}Holder, G.P., Mohr, J.J., Carlstrom, J.E., Evrard,
A.E., \& Leitch, E.M. 2000, ApJ, 544, 629

\reference{}Inogamov, N.A., \& Sunyaev, R.A. 2003, AstL, 29, 791

\reference{}Jones, C., \& Forman, W. 1984, ApJ, 276, 38

\reference{}Jones, M.E. 2002, ASP Conf. Ser. 257, \textit{AMiBA
2001: High-$z$ Clusters, Missing Baryons, and CMB Polarization},
eds. Lin-Wen Chen, Chung-Pei Ma, Kin-Wang Ng, and Ue-Li Pen (San
Francisco: ASP), 35

\reference{}Kaiser, N. 1986, MNRAS, 222, 323

\reference{}Kaiser, N. 1991, ApJ, 383, 104

\reference{}Kim, W.-T., \& Narayan, R. 2003, ApJ, 596, L139

\reference{}King, I.R. 1972, ApJ, 174, L123

\reference{}Kitayama, T., et al. 2004, PASJ, 56, 17

\reference{}Komatsu, E., et al. 2001, PASJ, 53, 57

\reference{}Kosowsky, A. 2003, NewAR, 47, 939

\reference{}Lamagna, L. et al. 2005, in \textit{Background Microwave
Radiation and Intracluster Cosmology}, ed. F. Melchiorri and Y.
Rephaeli, in press

\reference{}Lapi, A., Cavaliere, A., \& De Zotti, G. 2003, ApJ,
597, L93

\reference{}Lapi, A., Cavaliere, A. \& Menci, N. 2005, ApJ, 619, 60

\reference{}Lo, K.Y. 2002, in ASP Conf. Ser. 257, \textit{AMiBA
2001: High-$z$ Clusters, Missing Baryons, and CMB Polarization},
eds. Lin-Wen Chen, Chung-Pei Ma, Kin-Wang Ng, and Ue-Li Pen (San
Francisco: ASP), 3

\reference{}Lombardi, M., et al. 2005, ApJ, in press (preprint
astro-ph/0501150)

\reference{}Lucy, L. B. 1974, AJ, 79, 745

\reference{}Mahdavi, A., Boehringer, H., Geller, M.J., \& Ramella,
M. 1997, ApJ, 483, 68

\reference{}Markevitch, M. 1998, ApJ, 504, 27

\reference{}Masi, S., et al. 2003, Mem. SAIt, 74, 96

\reference{}Mazzotta, P., Brunetti, G., Giacintucci, S., Venturi,
T., \& Bardelli, S. 2004a, J. Korean Astron. Soc., 37, 381

\reference{}Mazzotta, P., Rasia, E., Moscardini, L., \& Tormen, G.
2004b, MNRAS, 354, 10

\reference{}McCarthy, I.G., Babul, A., Holder, G.P., \& Balogh, M.L.
2003, ApJ, 591, 515

\reference{}McCarthy, I.G., Balogh, M.L., Babul, A., Poole, G.B., \&
Horner, D.J. 2004, ApJ, 613, 811

\reference{}McNamara, B.R., et al. 2005, Nature, 433, 45

\reference{}Molendi, S., \& Pizzolato, F. 2001, ApJ, 560, 194

\reference{}Mushotzky, R.F. 2004, in \textit{Clusters of Galaxies:
Probes of Cosmological Structure and Galaxy Evolution,} ed. J.S.
Mulchaey, A. Dressler, and A. Oemler (Cambridge: Cambridge Univ.
Press), 124

\reference{}Narayan, R., \& Medvedev, M.V. 2001, ApJ, 562, L129

\reference{}Navarro, J.F., Frenk, C.S., \& White, S.D.M. 1997, ApJ,
490, 493

\reference{}Norman, M. 2005, in \textit{Background Microwave
Radiation and Intracluster Cosmology}, ed. F. Melchiorri and Y.
Rephaeli, in press

\reference{}Osmond, J.P.F., \& Ponman, T.J. 2004, MNRAS, 350, 1511

\reference{}Peebles, P.J.E. 1993, \textit{Principles of Physical
Cosmology} (Princeton: Princeton Univ. Press)

\reference{}Piffaretti, R., Jetzer, Ph., Kaastra, J.S., \& Tamura,
T. 2005, A\&A, 433, 101

\reference{}Ponman, T.J., Sanderson, A.J.R., \& Finoguenov, A.
2003, MNRAS, 343, 331

\reference{}Pratt, G.W., \& Arnaud, M., 2003, A\&A, 408, 1

\reference{}Reese, E.D., Carlstrom, J.E., Joy, M., Mohr, J.J.,
Grego, L., \& Holzapfel, W.L. 2002, ApJ, 581, 53

\reference{}Rephaeli, Y. 1995, ARAA, 33, 541

\reference{}Rephaeli, Y. 2005, in \textit{Background Microwave
Radiation and Intracluster Cosmology}, ed. F. Melchiorri and Y.
Rephaeli, in press

\reference{}Ricker, P.M., \& Sarazin, C.L. 2001, ApJ, 561, 621

\reference{}Roussel, H., Sadat, R., \& Blanchard, A. 2000, A\&A,
361, 429

\reference{}Sanderson, A.J.R., \& Ponman, T.J. 2003, MNRAS, 345, 1241

\reference{}Sarazin, C.L. 1988, \textit{X-Ray Emissions from
Clusters of Galaxies} (Cambridge: Cambridge Univ. Press)

\reference{}Schmidt, R.W., Allen, S.W., \& Fabian, A.C. 2004, MNRAS,
352, 1413

\reference{}Schwan, D., et al. 2003, NewAR, 47, 933

\reference{}Silk, J., \& White, S.D.M. 1978, ApJ, 226, L103

\reference{}Sunyaev, R.A., \& Zel'dovich, Ya.B. 1972, Comm. Astroph.
Sp. Sc., 4, 173

\reference{}Sutherland, R.S., \& Dopita, M.A. 1993, ApJS, 88, 253

\reference{}Tozzi, P., \& Norman, C. 2001, ApJ, 546, 63

\reference{}Valageas, P., \& Silk, S. 1999, A\&A, 350, 725

\reference{}Vikhlinin, A., Markevitch, M., Murray, S.S., Jones, C.,
Forman, W., \& Van Speybroeck 2005, ApJ, in press (preprint
astro-ph/0412306)

\reference{}Voit, G.M., \& Bryan, G.L. 2001, Nature, 414, 425

\reference{}Voit, G.M., Balogh, M.L., Bower, R.G., Lacey, C.G., \&
Bryan, G.L. 2003, ApJ, 593, 272

\reference{}Woody, D.P., et al. 2004, SPIE, 5498, 30

\reference{}Wu, K.K.S., Fabian, A.C., \& Nulsen, P.E.J. 2000,
MNRAS, 318, 889

\reference{}Yoshikawa, K., \& Suto, Y. 1999, ApJ, 513, 549

\reference{}Zaroubi, S., Squires, G., de Gasperis, G., Evrard, A.
E., Hoffman, Y., \& Silk, J. 2001, ApJ, 561, 600

\reference{}Zhang, T., \& Wu, X. 2000, ApJ, 545, 141

\end{references}
\end{document}